\documentclass[twocolumn]{pasj00}

\SetRunningHead{Astronomical Society of Japan}{Broad-Band Spectrum of The Black Hole Candidate IGR~J17497--2821 Studied with \emph{Suzaku}}

\title{Broad-Band Spectrum of The Black Hole Candidate IGR~J17497--2821 Studied with \emph{Suzaku}}

\author{Adamantia \textsc{Paizis}\altaffilmark{1,2}, Ken \textsc{Ebisawa}\altaffilmark{2}, Hiromitsu \textsc{Takahashi}\altaffilmark{3}, 
 Tadayasu \textsc{Dotani}\altaffilmark{2}, \\
 Takayoshi \textsc{Kohmura}\altaffilmark{4}, Motohide \textsc{Kokubun}\altaffilmark{2}, J\'er\^ome \textsc{Rodriguez}\altaffilmark{5}, 
 Yoshihiro \textsc{Ueda}\altaffilmark{6},\\
 Roland \textsc{Walter}\altaffilmark{7},  
 Shin'ya \textsc{Yamada}\altaffilmark{8}, Kazutaka \textsc{Yamaoka}\altaffilmark{9}, Takayuki \textsc{Yuasa}\altaffilmark{8}}

\altaffiltext{1}{Istituto Nazionale di Astrofisica, IASF\\
 Via Bassini 15, 20133 Milano, Italy}
\altaffiltext{2}{Institute of Space and Astronautical Science, JAXA\\
   3--1--1, Yoshinodai, Sagamihara, Kanagawa,  229-8510, Japan}
\altaffiltext{3}{Department of Physical Science, School of Science, Hiroshima University\\
1-3-1 Kagamiyama, Higashi-Hiroshima, Hiroshima 739-8526, Japan}
\altaffiltext{4}{Department of Physics, Kogakuin University\\
2665-1, Nakano-cho, Hachioji, Tokyo, 192-0015,  Japan}
\altaffiltext{5}{CEA Saclay, DSM/DAPNIA/Service d'Astrophysique\\
 FR-91 191 Gif-sur-Yvette Cedex, France}
 \altaffiltext{6}{Department of Astronomy, Kyoto University\\
    Kitashirakawa-Oiwake-cho, Sakyo-ku, Kyoto, 606-8502, Japan} 
\altaffiltext{7}{INTEGRAL Science Data Centre\\
Chemin d'Ecogia 16, 1290 Versoix, Switzerland}
\altaffiltext{8}{Department of Physics, University of Tokyo\\
   7-3-1, Hongo, Bunkyo-ku, Tokyo, 113-0033, Japan}
\altaffiltext{9}{Department of Physics, Aoyama Gakuin University\\
 5-10-1 Fuchinobe, Sagamihara, Kanagawa 229-8558, Japan}

\email{ada@iasf-milano.inaf.it}


\Received{$\langle$2008 July 15$\rangle$}
\Accepted{$\langle$2008 October 12 $\rangle$}
\Published{$\langle$publication date$\rangle$}
\def\chiq{{$\chi^2_{\nu}$}}

\def \compps {{\sc compps}}

\def \diskbb {{\sc diskbb}}
\def \cpl {{\sc cutoffpl}}
\def \wabs {{\sc wabs}}
\def \edge {{\sc edge}}

\def \nh {N$_{\rm H}$}
\def \xspec{{\sc xspec}}

\def \geom{{\sc geom}}

\def \kte{kT$_{\rm e}$}

\def \comptt{{\sc comptt}}
\def \gauss{{\sc gauss}}

\def \compps{{\sc compps}}

\def \relrefl{{\sc rel\_refl}}

\def\pl{{\sc pl}}

\def\cm2{cm$^{-2}$}
\def\s1{s$^{-1}$}

\def\wabs{{\sc wabs}}

\def\rxte{{\it RXTE}}
\def\swift{{\it Swift}}
\def\integral{{\it INTEGRAL}}
\def\chandra{{\it Chandra}}
\def\suzaku{{\it Suzaku}}

\def\igr{\mbox{IGR~J17497--2821}}

\begin{document}
\maketitle

\begin{abstract}
The broad-band 1--300\,keV \suzaku\ spectrum of \igr, the X-ray transient discovered 
by \integral\ in September 2006, is presented.  
\suzaku\  observed \igr\ on September 25,  eight days after its discovery, for a net exposure of about 
53\,ksec. During the \suzaku\ observation, \igr\ is very bright,  $2 \times 10^{37}$ erg s$^{-1}$  
at 8\,kpc in the 1--300\,keV range, 
and shows a  hard spectrum, typical of black hole candidates in the 
low-hard state. 
 Despite the multi-mission X-ray monitoring of the source, 
only with \suzaku\  is it possible to obtain a broad-band spectrum in the 1--300\,keV range with  a 
very high signal to noise
ratio. 
A sum of a multi-color disc (\diskbb) and a thermal Comptonization  component (\compps) 
with mild reflection is a good 
representation of our \igr\ \suzaku\ spectrum. The spectral properties of the 
accretion disc as well as the cut-off energy in the spectrum at about 150\,keV are clearly  detected and constrained.
We discuss the implications on the physical model used to interpret the data and the comparison with previous results.

\end{abstract}

\section{Introduction}
On 2006 September 17 a new hard X-ray transient, \igr\ \citep{soldi06}, was discovered by \integral\ \citep{winkler03}.
In the quest to understand the nature of this new transient, many follow-up observations were performed. Neither 
X-ray bursts nor pulsations, that would point to the presence of a neutron star, were found in \rxte\ data \citep{rodriguez07}.
\integral-\swift\  \citep{walter07} and \rxte\ \citep{rodriguez07} studies showed that the source had a very hard spectrum
 being  detected up to about 200\,keV. A phenomenological description of the \integral-\swift\ data 
 led to an absorbed cut-off power-law with photon spectral index $\Gamma$$\sim$1.7 and cut-off 
 around 200\,keV \citep{walter07}. 
The spectrum was also 
fitted with the thermal Comptonization model \comptt\ (Titarchuk 1994) obtaining a  plasma temperature of about 35\,keV. As the authors
point out, this temperature is coupled to the optical depth of the plasma and 
the  resulting temperature was poorly constrained: kT$_{\rm e}$=35$^{-9}_{+200}$\,keV.\\
Also Rodriguez et al., (2007) fit the data  with \comptt\ but, as they also point out, the 
parameters inferred were limited by the energy range studied (3--180\,keV): 
the seed photon temperature  was frozen to 0.1\,keV
and a Comptonizing plasma temperature of about 35\,keV and optical depth $\tau\sim$2 was obtained.
Two weeks after the source discovery, a \chandra\ observation was performed \citep{paizis07}. The narrow 
 \chandra\ energy range was not sensitive to the high energy cut-off and the spectrum could be fit with an absorbed 
power-law with a slope of about 1.2. \\
 Optical, near infra-red (NIR) and radio counterparts were searched for thanks to the \chandra\ 
  0.6 arcsec positional accuracy. 
The NIR counterpart was identified with a red giant K-type companion \citep{paizis07,torres06}. The NIR variability of the source 
\citep{torres07,walter07}  confirmed the  
association and was
consistent with a late-type companion located close to the Galactic center. Only an upper 
limit could be found for the
 optical band (V$>$23), as expected
for a source placed near or beyond the Galactic centre.
No radio counterpart was found down to a limit of 0.21\,mJy at 4.80 and 8.64\,GHz \citep{rodriguez07}.  Consistently with 
 previous studies, we assume an 8\,kpc distance to the source.\\
As can be seen, \igr\ has been throughly observed  during its outburst decay and the information gathered from 
various X-ray, radio and NIR 
observations, points to the source being  an X-ray Nova, a Low Mass X-ray Binary (LMXB) hosting a black hole (BH).\\
About eight days after the source discovery, \suzaku\ observed \igr\ with a net exposure of about 53\,ksec. \igr\ was at 80\% 
of its peak flux, about 80\,mCrab in the 15--50\,keV band. Preliminary results
were presented by Itoh et al. (2006) and showed that the source could be detected up to  $\sim$300\,keV. 
In this work we analyze in detail the \suzaku\ observation of \igr.

\section{Observation and Data Analysis}
\suzaku\ \citep{Mitsuda06} observed \igr\ about eight days after its discovery, starting from 2006 
September 25, 07:00 UT, up to September 26, 14:14 UT, for a total of about 53\,ksecs.
The data were acquired with
the X-ray Imaging Spectrometer (XIS, 0.12-12\,keV: \cite{Koyama06}) 
and the Hard X-ray Detector (HXD, 10--600\,keV: \cite{hxd1,hxd2}). The XIS detectors are located at the aim-points of the XRT mirrors 
\citep{Serlemitsos06}.  The viewing of the current observation is HXD-nominal.

 \subsection{XIS Data Selection and Processing}
 \label{section:XIS}
Given the brightness of  \igr,  XIS was mainly operated with the Burst and Window (1/4)
options, with only a few tens of seconds of data taken in the Normal mode (ignored in this work). 
These options, more suitable for bright sources in order to minimize pile-up,  
prevented us from carrying on power spectra analysis since about 50\% of the 2\,sec GTIs 
are data-free.\\
 Out of the four XIS, the three front-illuminated ones (FI, XIS 0 2 3) were mostly used in the 2x2 editing mode with 
only a small fraction of data taken in the 3x3 mode, while the fourth XIS (the back-illuminated, BI CCD, XIS 1)
was operated only in 3x3 editing mode. 
During the observation a dark frame/light leak error occurred in the sensors almost simultaneously,
the error being restricted to RAWY$<$118. The inclusion of these data resulted in an apparent flux 
decrease in all XIS sensors hence in all the analysis the data from the segments with RAWY$<$256 were not used.
This selection results in a $\sim$25\% reduction in the XIS normalization constant. 
Partial telemetry saturation occurring in one of the four segments reduces a few percent 
more the XIS normalization. Apart from the dark frame correction, the data were analyzed in the standard 
way. 

 \subsection{HXD Data Selection and Processing}
 \label{section:HXD}
\begin{figure}
\centering
\includegraphics[width=0.9\linewidth]{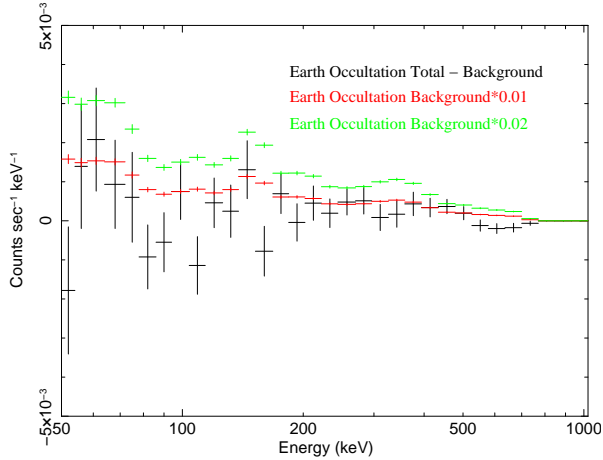}
\caption{Earth occultation total-background spectrum (black) used to compare between the data and the model prediction. 
The 1\% (red) and 2\% (green) levels of the background are also shown.}
\label{fig:back}
\end{figure}

The HXD was operated in the normal mode and we performed the analysis in the 
standard way described in the \suzaku\ data reduction documentation\footnote{http://www.astro.isas.ac.jp/suzaku/analysis/}.
Unlike XIS, HXD is not an imaging instrument and it is not possible to obtain background
data from the observation data themselves. Hence the HXD team has developed a model of the time-variable
particle background (Non X-ray Background, NXB). HXD NXB and response files 
 were downloaded from the instrument team web pages \footnote{PIN: http://www.astro.isas.ac.jp/suzaku/analysis/hxd/pinnxb/ GSO: http://www.astro.isas.ac.jp/suzaku/analysis/hxd/gsonxb/}, together with the GSO
 additional ancillary file distributed by the instrument team to match the Crab 
spectrum in the GSO energy range\footnote{http://www.astro.isas.ac.jp/suzaku/analysis/hxd/gsoarf/}. 
Given the source brightness, the cosmic X-ray background spectrum was not 
 included in the 
HXD analysis (less than 0.6\% of the observed rate). \\
A limiting factor for the sensitivity of the HXD-GSO is the reproducibility in the 
background estimation (systematic error), rather than the statistical error. 
In order to  accurately assess the systematic error specific to our  observation, 
we compared the nominal NXB spectrum (the model) with the spectrum obtained during earth occultation, 
as done in  \cite{fukazawa08}. 
The earth 
occultation 
``total-background'' spectrum is shown in black in Fig.~\ref{fig:back}. In the ideal case 
(model=real background) we would expect 
the subtracted spectrum to be equal to zero, whereas in the plot it is slightly over zero and close to
 the red spectrum (1\% of the background spectrum). This means that during the 
 earth occultation 
time of our observation, and presumably also during the nearby source observation time, the NXB model 
spectrum is underestimated 
and is most likely about 1\% higher. 
In our analysis we have applied a 
1.2\%  increase in the background spectrum that  corresponds to a 2$\sigma$ deviation 
in the all-data distribution of \cite{fukazawa08}.\\
We note that while the  HXD field of view below 100\,keV is very small,
34$^{\prime}$$\times$34$^{\prime}$, 
minimizing source contamination issues, the field of view above 100\,keV is 
of 4.5$^{\circ}$$\times$4.5$^{\circ}$, hence contamination could be occurring. 
During the \igr\ outburst, \integral\ detected only two sources above 100\,keV: \igr\ and 
1E~1740.7--2942 (the "great annihilator"), 2$^{\circ}$ away \citep{walter07}, with \igr\ clearly out-shining the "great annihilator"  at its outburst peak.
The \suzaku\ observation occurred when \igr\ was still at 80\% of its peak flux and very little contamination 
is expected to occur between 100--160\,keV where the transmission rate of the collimator is as low as 10\% for a 
2$^{\circ}$ off axis source in the HXD field. No contamination  is expected above 160\,keV where
\igr\ is the only active source, see Fig.~1 in 
Walter et al., 2007.

\begin{figure}
\centering
\includegraphics[width=0.69\linewidth,angle=270]{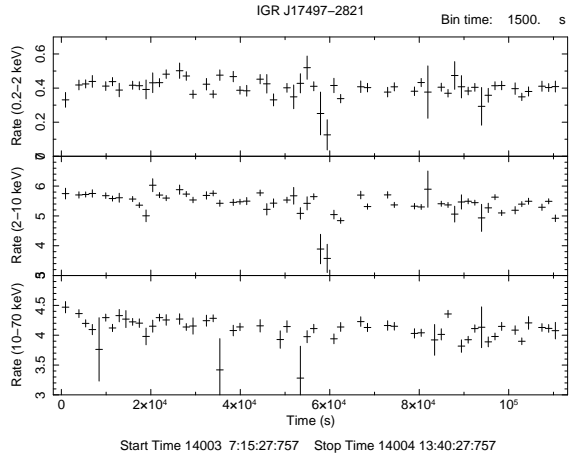}
\caption{BI-XIS and PIN background subtracted \igr\ lightcurves with a 1500\,sec time bin are shown. 
From top, the three panels refer to soft XIS band (0.2--2\,keV), hard XIS
(2--10\,keV) and PIN  (10--70\,keV). 
\label{fig:lc}}
\end{figure}

\subsection{Spectral fitting}
\label{section:xspec}
  Spectral fitting was performed with the \xspec\ v12 software package.
   The XIS source spectra obtained
 with the two different editing modes (2x2 and 3x3) in XIS 0 2 3 have been co-added in a final "XIS-FI-spectrum",  
 whereas the "XIS-BI-spectrum" has been kept separate
due to the differences in the  response matrix.
In the spectral fitting the 1--8\,keV XIS range was used, allowing for an  energy 
offset 
within $\pm$17\,eV, to account  for the current uncertainty in the XIS energy scale \citep{makishima08}. 
The gain of the two XIS spectra was also left free and always resulted to be close to 1 (1\% deviation in the 
worst case).
As for HXD, spectral fitting was carried on in the 12--70\,keV 
and 70--300\,keV band for PIN and GSO, respectively. \\
The overall model normalizations were allowed to differ between the XIS-FI, XIS-BI and HXD spectra but were 
constrained to be the same between HXD-PIN and HXD-GSO that we fixed  to  1.13. 
No systematic error was added to the XIS data.

\section{Results}
%
\label{res}
In this section we describe the results we obtain from the \suzaku\ data analysis both in terms of phenomenological 
and physical models. In the end, we include our results in the frame of a wider spectral evolution study 
of the \igr\ outburst, treating in a consistent way all the currently available high energy spectra of the source.

\subsection{\suzaku\ results: from phenomenology to physics}
In Fig.~\ref{fig:lc} the XIS (soft and hard X-ray band) and HXD-PIN count rate evolution of \igr\ is given 
(1500\,sec time bin). 
The overall XIS 2--10\,keV plot shows a slow but constant rate decrease along our observation. A similar
behaviour occurs also in the  PIN 10--70\,keV showing that the source did not switch  
between different spectral states. The XIS to PIN rate hardness ratio is quite constant 
hence we add all the available data to obtain one time-averaged spectrum per instrument.
This is a good representation of the source spectral behaviour and furthermore increases 
the statistics of the spectra, essential for the high energy cut-off investigation.\\
\begin{figure}
\centering
\includegraphics[width=0.65\linewidth,angle=270]{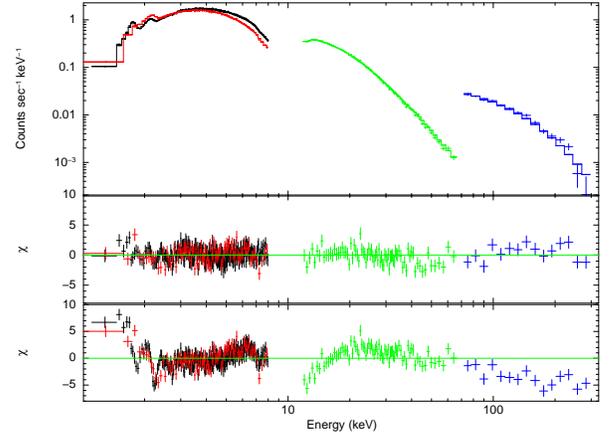}
\caption{
\emph{Upper panel} Background subtracted time-averaged XIS-FI (black), XIS-BI (red), HXD-PIN (green) 
and HXD-GSO (dark blue) spectra of \igr.
The \emph{intermediate panel} shows the residuals using the best fit \wabs*\edge*\cpl\ model given in Tab~1 
 while the lower panel shows the single \wabs*\pl\ fit model. See text.
\label{fig:step01}}
\end{figure}
To have the feeling of the shape of the spectrum, we first decided to find the appropriate phenomenological model to fit the 
data starting from the simplest case, a single absorbed power-law (\wabs*\pl, with absorption model by \cite{morrison83}).
This model is clearly not a good representation of the data as can be seen in Fig.~\ref{fig:step01}, \emph{lower panel}. 
Hence we preceeded
with an absorbed power-law with an exponential roll-over (\cpl). 
The fit improved significantly,
confirming that the overall \igr\ spectrum ($\Gamma$$\sim$1.45) indeed breaks above 80\,keV. 
Fig.~\ref{fig:step01} \emph{medium and upper panel} show the best fit obtained including also 
an absorption edge (\edge),  needed in the XIS data when investigated separately from 
HXD (probability of $\sim$10$^{-5}$ that the improvement was purely due to chance).
In the fit using only XIS data, the (fixed) 7.11\,keV Fe edge shows an optical depth of  0.05$\pm$0.02. 
Hereafter in the complete XIS-HXD fits performed, 
we freeze the value of the 2 parameters describing the absorption edge to what obtained from the XIS 
spectra alone. Details of the final best fit model are given in Table~1. 
A background deviation from the 1.2\% case used here (Section~\ref{section:HXD}) does not affect the power-law slope value 
that is mainly driven by the  XIS$+$PIN spectral shape. Indeed a  $\Gamma$=1.45$\pm$0.01 is obtained for the nominal, 
0\% increase, NXB.
\small
\begin{table}
\begin{center}
\caption{Best-fit model: \wabs*\edge*\cpl$^{\dag,\ddag}$.}
\begin{tabular}{lccccc}
\hline
\hline
\nh  &	$\Gamma$  & E$_{cut}$   & E$_{edge}$ & Tau$_{edge}$\\
($10^{22}$ cm$^{-2}$) & & (keV) & (keV) & \\
\hline
& & &  \\
 4.56$\pm$0.02 &  1.45$^{+0.02 }_{-0.01}$  & 150$^{+7 }_{-4}$  & [7.11] & [0.05] \\ 
& & &  \\
\multicolumn{2}{l}{\chiq(dof)=1.085 (1158)} \\
& & &  \\

\hline
\hline
 \end{tabular}
\end{center}
$^{\dag}$ Errors are computed at 90\% confidence level for a single parameter.\\
$^{\ddag}$ The parameters of the \edge\ component were frozen to the best fit values obtained from the XIS fit alone. \\

\label{tab:fit}
\end{table}
\normalsize

\begin{figure}
\centering
\includegraphics[width=0.65\linewidth,angle=270]{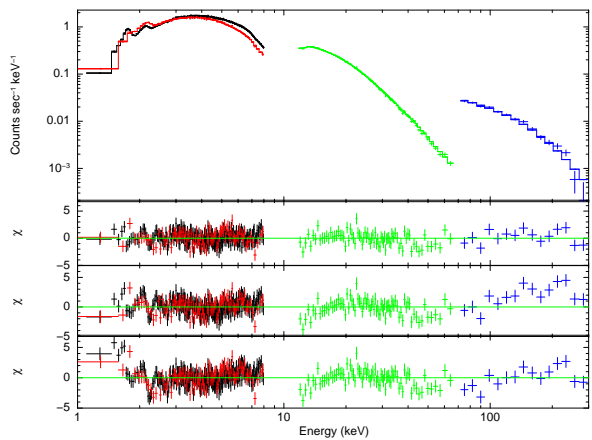}
\caption{As in Fig.~\ref{fig:step01} but with \wabs*\edge*\compps\ 
(residuals in the lower panel), \wabs*\edge*(\diskbb+\compps) 
(residuals in the middle panel), and  \wabs*(\gauss+\diskbb+\compps) with reflection 
(residuals in the upper panel and fit, Table~2).
\label{fig:step432}}
\end{figure}

\begin{figure}
\centering
\includegraphics[width=0.8\linewidth]{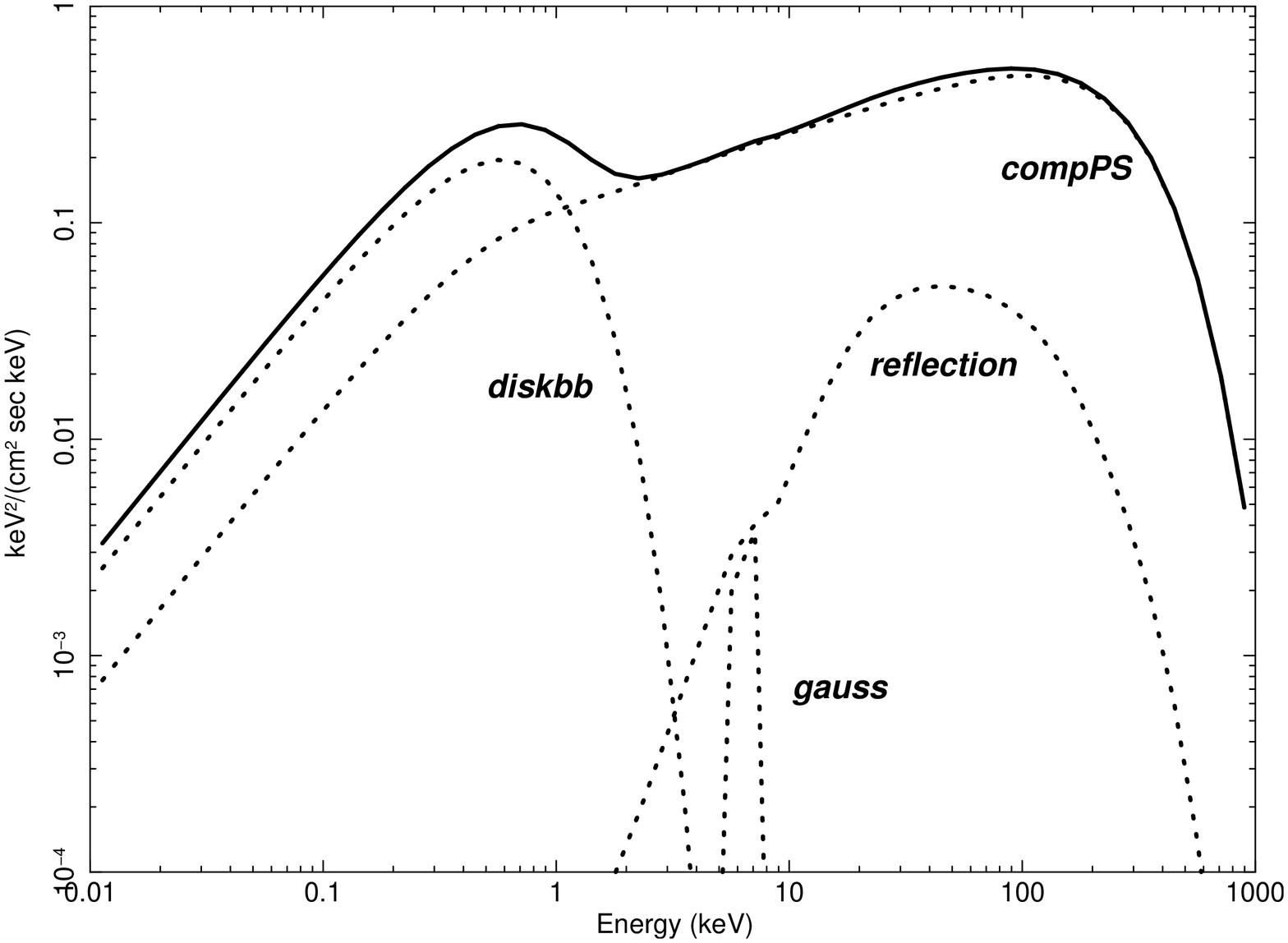}
\caption{Unabsorbed best fit model used to fit the \igr\ \suzaku\ spectrum, \gauss+\diskbb+\compps\ in Table~2..
\label{model}}
\end{figure}

A cut-off power-law shape suggests that the spectrum can be described in terms of a thermal
Comptonization model. Among all the available models we choose 
\compps\ \citep{Poutanen96}. 
The high energy cut-off found from the phenomenological fit 
implies a hot Comptonizing corona  for which \compps\ (unlike e.g.~\comptt)
is suitable. 
Fitting the \igr\ spectrum with an absorbed \edge*\compps\ component  did not 
result in a good fit 
as can be seen in Fig.~\ref{fig:step432}, \emph{lower residual panel}.
A  clear soft excess 
is visible in the low end of XIS data as well as a clear "S" shape in the HXD domain (\chiq=1.3, 1157 dof).
As a first improvement we add a soft component to account for the excess. A natural candidate
is a cold disc emission and we choose the \diskbb\ representation of it \citep{mitsuda84}. 
The seed photon temperature of the two models (\diskbb\ and \compps) were constrained to a single value and the model 
used is an absorbed \edge*(\compps+\diskbb).
Fig.~\ref{fig:step432}, \emph{intermediate residual panel}, shows that although the soft excess description improved, 
the HXD part of the spectrum is still not correctly described ("S" shape in the residuals, \chiq$\sim$1.2, 1156 dof). \\
The residuals of the high energy detector, ``S'' shape, resemble what obtained by \cite{makishima08}  and \cite{hiro08} 
in the study of two other BH binaries, Cyg~X--1 and \mbox{GRS~J1655--40}, respectively. 
Similarly to these authors, we included the reflection option 
in the \compps\ model, letting free the \relrefl\ parameter that indicates the solid angle of the cold material 
visible
from the Comptonizing source, in units of 2$\pi$. Since the reflection component includes 
also the  absorption edge, we  did not include it explicitely in the model.  Furthermore, 
in the spectral fitting,  
we included a 6.4\,keV Fe line, expected in the presence of reflection from a cold disk. In the broad band 
fit, the line width was frozen to the 
best fit value obtained by the spectral fitting of XIS data alone, 0.39\,keV.
With a reflection of 
about 0.17  the fit improved significantly (\chiq$\sim$1.07, 1154 dof) as shown in Fig.~\ref{fig:step432}, 
\emph{upper residual panel and fit}. Details of this best fit model are  shown in  
Table~2 while the unabsorbed best fit model is shown in Fig.~\ref{model}. \\
In the \compps\ model we have assumed a spherically symmetric corona (\geom=4, see Table~2)
where the seed photons are injected in the center of the sphere. 
Under such a configuration, the energy spectrum of the Compton up-scattered
emission has no dependence on the source inclination angle, that we assumed to be \emph{i}=$60^\circ$.\\
We note that unlike the case of \cite{makishima08}  and \cite{hiro08} a two component \compps\ model  besides \diskbb\ is not needed.
While in their case this interpretation was supported by the fact that the two different 
Comptonizing regions indeed showed different optical depth (ratio of a few), in our case this is not true
and a double \compps\ model still returns a single Comptonizing region. 
The reason for this could be the high absorption we detect in the spectrum of \igr\ with respect to GRO~J1655--40 and Cyg~X--1 (an order 
of magnitude higher). This, combined to the worse statistics, hampers the detection of the second "soft" \compps\ that dominated 
below $\sim$3\,keV  (compare our unabsorbed best fit model shown in Fig.~\ref{model}, with Fig.~5 in \cite{makishima08}). \\
 We believe that the model given in Table~2 
is the best physical representation we can extract from the current data, \chiq\ is acceptable and 
there is no clear feature in the obtained residuals
that would require a physics-driven additional component.
\small
\begin{table*}
\begin{center}
\caption{
Best-fit model:\wabs*(\gauss+\diskbb+\compps)$^{\S,\dag,\S\S}$

}
\begin{tabular}{cccccccccc}
\hline
\hline
     \nh  			    &	kT$_{\rm in}^{\rm diskbb}$    &r$_{\rm in}^{\rm diskbb}$ &kT$_{\rm in}^{\rm compps}$      &r$_{\rm in}^{\rm compps}$   &\kte&  $\tau$ & Refl. &  EW$_{line}$$^{\P}$\\
     ($\times 10^{22}$ cm$^{-2}$) &  (keV)  & (km) &  (keV)       & (km) & (keV)    &   &&   (eV)   \\
\hline
  & & &  & & & & &\\
 5.15$^{+0.01 }_{-0.03 }$  & 0.25$^{+0.01 }_{-0.02 }$ & 123$^{+33}_{-4}$& =\diskbb& 53$^{+2}_{-4}$$^{\ddag}$  & 85$\pm$9 & 1.42$^{+0.25 }_{-0.15}$  & 0.17$\pm$0.03& 37$\pm$10 \\
  & & & & & & & &\\
\multicolumn{2}{l}{\chiq(dof)=1.07 (1154)} \\
  & & & & & & & &\\
 \hline
 \hline

 \end{tabular}
\end{center}
$^{\S}$ Errors are computed at 90\% confidence level for a single parameter.\\
$^{\dag}$ \igr\ is assumed to be at a distance of 8\,kpc with an inclination \emph{i}=$60^\circ$.\\
$^{\S\S}$ A spherical corona configuration is assumed i.e. \geom=4.\\
$^{\ddag}$ Not corrected for the inclination, see text. \\
$^{\P}$ The center and width of the line were fixed at 6.4\,keV and 0.39\,keV respectively, as obtained by the XIS data alone. See text.
\label{tab:physic}
\end{table*}

\normalsize

\subsection{Long-term spectral variability}
\begin{figure}
\centering
\includegraphics[width=0.9\linewidth]{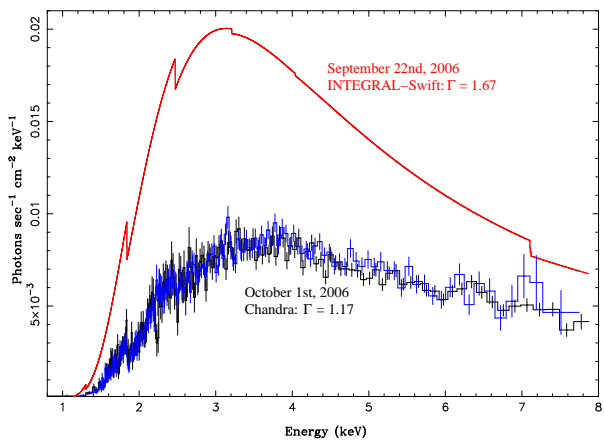}
\caption{Best fit model of the outburst peak compared to the  Chandra follow-up data. Axes have been left linear to 
better visualize the slope differences. 
\label{fig:compare}}
\end{figure}

\igr\ has been  observed by more than one high energy mission along its decay.
Different missions introduce instrumental differences, furthermore it is not always possible to compare the results 
published by the authors
due to different models used. To overcome the bias introduced from different models, we
decided to fit all the published spectra in terms of the same model, to ease the comparison\footnote{Co-authors of this 
paper have been the leading authors of the multi-mission follow-up.}. We choose the model 
of Table~1: \wabs*\edge*\cpl. In all cases the values of \nh\ and the \edge\  were frozen to the values of Table~1, 
 to avoid the bias they introduce in the power-law 
photon index. Only 
three parameters, $\Gamma$, E$_{cut}$ and normalization, were let free.
\rxte\ data, presented in \cite{rodriguez07}, span from September 20 up to September 29, 2006. During all the \rxte\ follow-up, 
\igr\ remained constant with $\Gamma\sim$1.55. \integral-\swift\  \citep{walter07} and \suzaku\ spectra (this work)
were taken during the period covered by \rxte\ (20-22 September the former and 25-26 September the latter). Our fit  
returns a power-law slope $\Gamma$=1.67$\pm$0.05 for the former case (\integral-\swift), and $\Gamma$=1.45$^{+0.02 }_{-0.01}$ in  the latter
 (\suzaku, Table~1).
 Most likely this slope difference is due to instrumental cross-calibration effects since it occurs during the period when the source
 was seen to be constant by \rxte.
The \chandra/HETGS spectrum \citep{paizis07}  that refers to a later observation, October 1, returns a slope 
of $\Gamma$=1.17$\pm$0.02, considerably harder than what previously obtained.  In this case we have frozen 
E$_{cut}$ to the \suzaku\ value, 150\,keV, being the cut-off considerably beyond the \chandra\ energy range.
The source is rather 
constant during the \chandra\ observation and instrumental issues, at the origin of the hard spectrum, such as a high pile-up 
are very unlikely since the pile-up fraction in the HETGS arms 
is about 2\% \citep{paizis07}.  
A possible reason for the hard spectrum obtained in \chandra\ data could be the heavy absorption of the system that leaves a \chandra~effective
energy range that is very narrow (1.5-8\,keV), possibly introducing a bias in the result.
To test this hypothesis, we fit the \suzaku/XIS data alone so that we "reproduce" the \chandra\ energy range. 
Fixing \nh, \edge\ parameters and E$_{cut}$ to what found in Table~1, we find that the XIS spectrum is still 
consistent with the broad band result, $\Gamma$=1.45$\pm$ 0.01 (\chiq=1.042, 1064 dof) hence the 
\chandra\ spectrum is indeed hard. This suggests that \igr\ remained 
roughly constant for about 9 days (\rxte\ period) to suddenly harden two days later. 
Another possibility is that a change of hydrogen column density occurred in the system 
causing the apparent change of slope. Indeed \igr\ appears to have a significant contribution from the local matter 
near the binary system, its hydrogen column density \nh=(4--5)$\times 10^{22}$ cm$^{-2}$  is in fact larger that the 
galactic one $\sim$1.5$\times 10^{22}$ cm$^{-2}$ \citep{dickey90}. Nevertheless, the spectral change cannot be ascribed 
to a variability of \nh\ alone since the \chandra\ spectrum is not compatible with a fixed \suzaku\ value of $\Gamma$=1.45, no matter what the 
value of \nh\ (\chiq=1.32 for 336 dof and probability of $\sim$10$^{-13}$ that the improvement was purely due 
to chance from a frozen $\Gamma$=[1.45] to a free $\Gamma\sim$1.2).
   Hence, most likely, a combination of an increase in \nh\ and spectral hardening occurred along the outburst.\\
We note that the attempt to explain the spectral hardening in terms of the disappearing disk (\diskbb) 
alone does not work since the \suzaku\ data require the \emph{shape} of the spectrum to be changed and not the 
normalizations of the two components alone.\\
Fig~\ref{fig:compare} shows the two extremes: the  \integral-\swift\ peak best fit model, together with the \chandra/HETG spectra.
In about nine days (from peak flux to \chandra\ follow-up), the 2--10\,keV flux of \igr\ decreased by about 50\% whereas the 
15--50\,keV flux decreased by about 20\% only \citep{paizis07}, i.e. the source seems to have started hardening  
about nine days after the peak.

Besides \suzaku, only the \chandra\ observation of \igr\ \citep{paizis07}, 
did show evidence of an accretion disc (kT$_{\rm in}\sim$0.2\,keV) and it would be interesting to compare the temperature and radius of the disc in the two 
cases. Unfortunately the overall spectral shape is different in the two cases and in the soft part of the spectrum,
where \nh, \diskbb\ and soft end of the \pl\ co-exist, it is not possible to disentangle the different contributions to isolate 
the  disc variability.

\section{Discussion}
\label{discussion}
We have presented the broad band X-ray spectrum  of 
\igr\ obtained with \suzaku\ eight days after the discovery and investigated the long-term 
spectral variability in the two weeks following the source discovery.
During the \suzaku\ observation, the source unabsorbed flux in the 1--300\,keV band 
is about $3 \times 10^{-9}$ erg s$^{-1}$ cm$^{-2}$ that, for an assumed distance of 8\,kpc, 
 leads to a luminosity of about $2\times 10^{37}$ erg s$^{-1}$.

 \subsection{\suzaku\ broad-band spectrum and implications}
The high signal to noise ratio achieved in the \underline{soft end of the \suzaku\ spectrum}, together with the low energy 
threshold, allowed us to constrain the parameters that describe the thermal part of the spectrum.
Unlike what obtained with \integral\ and \rxte, the soft component, interpreted as the accretion disk, 
is well constrained, see Table~2. This is also the case for the reflection component  detected for the first 
time in \igr.
 The \diskbb\ component has a raw innermost radius r$_{in}$$\sim$123\,km (assumed  8\,kpc
 distance and $60^\circ$ inclination) which yields to the best-estimate actual inner radius R$_{in}$$\sim$145\,km,
 obtained applying nominal correction factors in \cite{makishima00}, including the color hardening factor of 1.7.
 Conversely, the \compps\ component has an innermost radius r$_{in}^{seed}$$\sim$53\,km that leads to
 a corrected inner radius R$_{in}^{seed}$$\sim$62\,km\footnote{R$_{in}^{seed}$ was not corrected for the inclination 
 since the Comptonized emission is more likely isotropic.}. 
These two inner radii provide an indication of the size of the emitting areas and in our case
  the region  directly visible is about (123/53)$^{2}$ = 5.4 times 
larger than the seed photon region (i.e. $\sim$84\% of photons are directly visible and 16\% 
are covered by the Compton cloud). 
This may suggest that the 
 Comptonizing plasma has a patchy structure, as suggested also by \cite{gierlinski97} and more recently by
 \cite{makishima00} in the case of Cyg~X--1.\\
In order to calculate the best estimate of the inner radius of the (optically thick) disk,
we need to measure its total flux.   In our case,
the normalization of the \diskbb\ model alone represents only a part of
the disc flux since a  fraction  is Compton
up-scattered by the corona and is not observed directly.
Hence to obtain the total disc flux it is necessary to combine the information 
from the \diskbb\ and \compps\ models, adding quadratically the two obtained inner radii,
 as done in e.g. \cite{makishima08}. We obtain a disc inner radius of 158\,km, 
 $\sim11R_{g}$ for a 10 solar mass BH.
This estimated value for the inner disc radius may  be relatively small 
compared to the idea of disc truncation 
in the LHS, but it should not be used to rule it out. Indeed, strong conclusions on the real inner disc radius cannot 
be driven since the results are subject to uncertainties such as  
the source inclination and distance assumed, the color hardening factor in the LHS, and the validity of 
the   multi-color optically thick description of the disc at all radii (see \cite{makishima08} for a detailed 
discussion).\\
At the \underline{higher end of the spectrum}, the cut-off energy we obtain is well below the upper energy boundary of the data, 
hence we are sampling both the rise and the descent of the spectrum.\\
 \igr\ is in a typical LHS in which  unsaturated Comptonization
is believed to be the main physical process at work. In this regime, a 150\,keV 
cut-off leads to an estimated 
corona temperature of about 75\,keV. Seed photons and Comptonizing plasma
are far from being in thermal equilibrium and a large fraction of energy is exchanged at each interaction.
In these conditions,  models such as \compps\ \citep{Poutanen96} that uses iterative
scattering numerical procedures is more suitable than e.g. \comptt. Indeed the latter model is basically valid in the diffusion regime, i.e. 
when the \emph{average} photon energy exchange per scattering is small ($\Delta E/E \ll 1$) and photons 
suffer many scatterings when traveling across the plasma before escaping  (see Farinelli et al, 2008, for a discussion on Comptonization models).\\
Our detection of the source up to 300\,keV strengthens the interpretation of an X-ray Nova:  Comptonization is 
believed to be the dominating process at work in these objects and a  signal up to 300\,keV implies 
very high plasma temperatures that are more common in  BHs in their LHS.  
Indeed, unlike for BHs, in the case of a NS the surface is an additional  source of seed photons 
that controls the plasma temperature through Compton cooling \citep{barret01}.

\subsection{X-ray Novae: far from canonical}
 Novae are known to show a series of different X-ray spectral states during their outburst.
The "canonical" evolution, i.e. the first one seen in X-ray transients, shows a spectral softening while the 
 flux increases and then a hardening  during decay, towards quiescence. The advent of many X-ray missions 
 together with their monitoring capabilities has proven that this standard evolution is far from being the only one. 
A certain number of X-ray Novae has shown outbursts with a single spectral state, 
 the Low Hard State with $\Gamma<$2 all along the outburst (e.g. Rodriguez et al., 2006, and references therein). 
 This is also the case for \igr\ that remained hard during the outburst, with a possible hardening along the decay.
 It has further been assumed that "canonical" X-ray transients follow the Fast Rise Exponential Decay morphology 
 whereas this is frequently not the case; even in the genuinely soft sources a precursor low/hard state phase is seen 
 prior to the fast rise in all cases where data are available \citep{brocksopp02}.
 Furthermore, a correlation between luminosity and spectral softness is not necessarily observed in all sources 
 \citep{brocksopp04} and  evidence for a spectral softening in the last phases of the decay 
 while entering in quiescence has also been found \citep{tomsick04}. 
 
 \section{Conclusions}
We have presented the best broad-band high energy spectrum currently available for the 
X-ray transient \igr. A fit with a simple phenomenological model gave a constrained 
high energy cut-off around 150\,keV, naturally driving the choice for the  model  used to 
interpret the data in a physical scenario. 
Soft seed photons coming from the disc are thermally up-scattered by a hot and patchy corona. A very mild reflection 
component is seen for the first time in the spectrum. Though not conclusive, there is a hint 
of spectral hardening about 10 days after the peak outburst with the source  remaining in its LHS 
all along the outburst.\\
Explaining within a self consistent scenario a "canonical" BH LMXB (e.g. XTE~J1550--564 \cite{rodriguez03}), 
the LHS outburst of e.g. \igr\ (this work) and the final softening towards quiescence (e.g. XTE~J1650--500 \cite{tomsick04}) 
is a real challenge, essential 
to really understand the physics of accretion. Broad-band studies are the first important step 
to have a complete view of the source spectral properties 
in the X-ray domain, where the accretion power resides.\\
 
 This work has been supported by the  Japan Society for the Promotion of Science (JSPS) grant
in  the frame of the short-term Postdoctoral Fellowship for Foreign Researchers (March-April 2008).
AP thanks JAXA/ISAS, where most of the work has been done, for their kind  hospitality 
and interesting scientific (and cultural!) discussions. AP acknowledges the Italian Space Agency financial and 
programmatic support via contract I/008/07/0.


\end{document}